\documentclass[aps,prd,a4paper,twocolumn,showpacs,showkeys,preprintnumbers,amsmath,amssymb]{revtex4}

\usepackage{graphicx}
\usepackage{bm}
\usepackage{dcolumn}
\usepackage{color}
\usepackage{pst-plot}
\renewcommand{\texttt}{{}}
\newcommand{\be}{\begin{eqnarray}}
\newcommand{\ee}{\end{eqnarray}}
\usepackage{graphicx}

\begin{document}

\title{Entropic force, noncommutative gravity and ungravity} 

\author{Piero Nicolini}
\email{nicolini@th.physik.uni-frankfurt.de}
\affiliation{Frankfurt Institute for Advanced Studies (FIAS),
Institut f\"ur Theoretische Physik, Johann Wolfgang Goethe-Universit\"at, 
Ruth-Moufang-Strasse 1, 60438 Frankfurt am Main, Germany}
\date{\today}
\date{\small\today}

\begin{abstract} \noindent
After recalling the basic concepts of gravity as an emergent phenomenon, we analyze the recent derivation of Newton's law in terms of entropic force proposed by Verlinde. By reviewing some points of the procedure, we extend it to the case of a generic quantum gravity entropic correction to get compelling deviations to the Newton's law. More specifically, we study: (1) noncommutative geometry deviations and (2) ungraviton corrections. As a special result in the noncommutative case, we find that the noncommutative character of the manifold would be equivalent to the temperature of a thermodynamic system. Therefore, in analogy to the zero temperature configuration, the description of spacetime in terms of a differential manifold could be obtained only asymptotically. Finally, we extend the Verlinde's derivation to a general case, which includes all possible effects, noncommutativity, ungravity, asymptotically safe gravity, electrostatic energy, and extra dimensions, showing that the procedure is solid versus such modifications.

\end{abstract}

\maketitle

\section{Introduction: emergent gravity} Entropy is one of the most intriguing concept in physics, which intersects the interest of researchers working in different sectors (see, for instance, \cite{classicalentropy}). Conventionally, entropy is defined as a measure of the loss of information about the microscopic degrees of freedom of a physical system, when describing it in terms of macroscopic variables. As a typical example to explain the concept of entropy, one deals with a gas or more in general with a fluid. At a microscopic level, the gas consists of a vast number of freely moving atoms or molecules. Microscopic degrees of freedom are described by the microstate of the system and refer to the positions  and momenta of all the atoms. In principle, all the physical properties of the system can be obtained from the microstate. However, provided the condition of thermodynamic equilibrium, the system can be adequately described by macroscopic quantities, such as volume, pressure, temperature, and so forth, which define the system macrostate. Kinetic theory provides an explanation about the connection between temperature and the molecular/atomic structure of matter. This connection affects also the concept of entropy. According to Boltzmann, entropy is 
\begin{equation}
    S = k_B \ln N,
\end{equation}
where $k_B$ is Boltzmann's constant and $N$ is the number of microstates compatible with the given macrostate. 

Entropy arises in  many different contexts, but, in particular, it can be seen as a consequence of the gravitational interaction. In 1975 Hawking showed, in terms of quantum particle pair production, the possibility for a black hole to evaporate, i.e., the existence of a thermal flux of radiation emitted from a black hole with a black body spectrum at temperature \cite{Hawking:1974sw} 
\begin{eqnarray}
T=\frac{\hbar c
\kappa}{2\pi k_{B}},  
\label{temp}
\end{eqnarray}
where $\kappa$ is the black hole surface gravity. As a consequence black hole entropy can be defined on the grounds of the concept of black hole evaporation. However the description of gravity in terms of general relativity, even if satisfactory in most aspects, is far from being complete. As far as the black hole emits radiation, it loses energy and shrinks. In such a way, the surface gravity $\kappa$ increases and the black hole becomes hotter and hotter. In the end the evaporation becomes a runaway phenomenon and the black hole background geometry develops a coordinate singularity. Therefore there are well-defined situations in which general relativity breaks down. In other words, in every situation in which an intrinsic microscopic scale, e.g. the Planck scale, is probed by a system, general relativity turns out to be an inadequate theory. This might suggest an analogy between the classical spacetime $g_{\mu\nu}(\textbf{x},t)$ and macroscopic quantities of a gas, since they both lose their meaning in the microscopic description. The classical spacetime $g_{\mu\nu}(\textbf{x},t)$ could be nothing but an emergent, low energy, large scale phenomenon of the spacetime microstructure, like gas macroscopic variables emerging from the molecular dynamics. As a consequence we could reverse our line of reasoning, saying that the gravitational interaction descends from spacetime thermodynamics (for instance see \cite{entropic}). Following \cite{paddy,paddyreview} we see that a metric
\begin{equation}
ds^2=-f(r)c^2dt^2+f^{-1}(r)dr^2+r^2d\Omega^2
\label{metric}
\end{equation}
is naturally associated to a temperature. If the function $f(r)$ admits a simple zero at $r=r_H$, we can expand $f(r)\approx f^\prime(r_H)(r-r_h)$ and cast the $r-t$ sector of  (\ref{metric}) in the form
\begin{equation}
ds^{2}_{2d}=-(\kappa x)^2dt^2+dx^2
\label{rindler}
\end{equation}
where the surface gravity is $\kappa=(c^2/2)f^\prime(r_H)$.  Equation (\ref{rindler}) is nothing but a Rindler metric. The Euclidean continuation of the time allows us to identify a temperature of the form (\ref{temp}). Entropy naturally emerges from the thermodynamic relation
\begin{equation}
k_B TdS=dU+PdV
\label{thermolaw}
\end{equation}
where $dU=(1/2G)c^4dr_H$, $PdV=Pd(4\pi r_H^3/3)$, and reads
\begin{equation}
dS=\frac{c^3}{G\hbar}d\left(\frac{1}{4}4\pi r_H^2\right).
\label{entropy}
\end{equation}
We note that, given the above definitions for the thermodynamic quantities,  we can derive from  (\ref{thermolaw}) the Einstein equation evaluated at the horizon $r=r_H$ for the line element in (\ref{metric}), i.e.,
\begin{equation}
(1-f)-rf^\prime=-\frac{8\pi G}{c^4}T_r^r r^2
\end{equation}
with $r=r_H$ and $P=T_r^r$. 
If, as it seems, gravity is an emergent phenomenon, we cannot forget that entropy encodes the nature of microscopic degrees of freedom of spacetime. Therefore, one might think to use entropy to link general relativity with the statistical description of the yet unknown spacetime microstructure.
On the other hand, all the above reasoning heavily depends on the presence of horizons. 
Against this background, in a recent paper, Verlinde overcame this stint \cite{Verlinde:2010hp}. 

\section{Entropic Newton's law} Verlinde's contribution attracted considerable media exposure \cite{media} and, though the paper is not yet published, he triggered the interest about this topic with immediate follow-up work in cosmology \cite{vcosmo}, black hole physics \cite{vblack}, loop quantum gravity \cite{vloop}, and other fields \cite{vother}. 

Verlinde's arguments are based on the fact that gravity is expected to be universal and independent of the details of the underlying spacetime microstructure. Further, he made new considerations about holography. As previously shown in \cite{holo},  generally covariant actions can be expressed in the form 
\begin{equation}
A=\int d^4x\sqrt{-g}R=\int d^4x\sqrt{-g}[L_{\mathrm{bulk}}+L_{\mathrm{sur}}].
\end{equation}
The surface term $L_{\mathrm{sur}}$ contains the information stored in horizons. However, the gravitational action shows holography in the sense that there is a relation between $L_{\mathrm{bulk}}$ and $L_{\mathrm{sur}}$, i.e.,
\begin{equation}
\sqrt{-g}L_{\mathrm{sur}}=-\partial_\mu\left(g_{\alpha\beta}\frac{\partial\sqrt{-g}L_{\mathrm{bulk}}}{\partial(\partial_\mu g_{\alpha\beta})}\right)
\end{equation}
which permits us to show that the same information is encoded in both the bulk and surface terms.  For this reason, the horizon entropy can be calculated by evaluating the surface term. As a result one finds an entropy proportional to the horizon area as in (\ref{entropy}), rather than to the volume inside the horizon, as one would expect in case of nongravitational entropies. Against this background, Verlinde assumes that holography in the presence of horizon is just a specific example of a much more general concept, which affects  Newton's law of gravitation too. For instance in the case of two masses, one test mass $m$ and a source mass $M$, we can consider a surface $\Omega$ centered around $M$and lying between the two masses. We also assume $m$ to be at a distance from $\Omega$ smaller with respect to its reduced Compton wave length $\lambda_m=\hbar/(mc)$. Given this point, Verlinde reversed the logic according to which the laws of gravity led to holography, assuming the following starting points: 
\begin{list}{}{}
\item[(1)] in the vicinity of the surface $\Omega$, the change of surface entropy is proportional to $\Delta x$, the change of the radial distance of the mass $m$ from the surface, i.e.,
\be
\Delta S_\Omega = 2\pi k_B \frac{\Delta x}{\lambda_m}
\ee
up to subleading terms ${\cal O}(\Delta x^2/\lambda^2_m)$. The proportionality constant is chosen to match $F=ma$, when the Unruh temperature for accelerated observers is considered;
\item[(2)] the energy of the surface is identified with 
\be 
U_\Omega = M c^2,
\ee
the rest mass of the source;
\item[(3)] on the surface $\Omega$, $N$ bits of information are stored, i.e., 
\be 
N=\frac{A_\Omega}{\ell_P^2}
\ee
where $A_\Omega$ is the area of $\Omega$ and $\ell_P$ is the Planck length;
\item[(4)]  the surface $\Omega$ is in thermal equilibrium at the temperature $T$, all bytes are equally likely and the energy of $\Omega$ is equipartitioned among them i.e., 
\be 
U_\Omega = \frac{1}{2} N k_B T;
\ee
\item[5)] a force $F$ arises from the generic form of the thermodynamic equation of state
\be
F\Delta x=T\Delta S_\Omega.
\ee
\end{list}
As a result, one can derive from the above assumptions the Newton's law $F=GMm/r^2$. 

Some comments are in order.
To obtain an entropic force, we just need to find a temperature.  Since, according to Unurh, temperatures are related to accelerations, a force descends. The novelty is that to have a temperature, Verlinde never mentions horizons. He just assumes that the surface $\Omega$ is an information storage device, i.e., an holographic surface. There is no assumption about the existence of the gravitational force. Gravity emerges through the information bit unit, namely $\ell_P^2$. This lets us define the gravitational constant as $G\equiv c^3 \ell_P^2/\hbar$. 

In support of these arguments, there is the expected universal character of gravity. If the gravitational force is entropic in the presence of horizons, the universality would let us interpret gravity as an entropic force even in the absence of horizons. More specifically the Newton's potential $\phi$ enters the line element of black hole solutions. Therefore if gravity is entropic, this would be independent of the specific value $\phi=-c^2/2$ for which a horizon occurs. Furthermore, by thinking about accelerated observers, one can have holographic screens anywhere in space. 
On the other hand Verlinde's work reception includes several criticisms too. The basic assumptions seem a way to justify dimensional manipulations, from which the entropic force equals the Newton's force as a mere coincidence. For instance in the derivation, one sets the area of the surface equal to $4\pi r^2$, where $r$ is later identified with the radial distance of the mass $m$ from the source $M$. Rather than embarking in further analyses about the robustness of the derivation, we prefer to study another matter. Since in Verlinde's derivation very little is devoted to possible improved versions of the entropy of the holographic surface, we intend to take into account effects due to quantum gravity/microscopic scale effects for the following purposes:
\begin{itemize}
\item[(a)]to extend the validity of the Verlinde's procedure taking into account improved forms of the entropy (\textit{universality}); 
\item[(b)] to derive deviations to Newton's law emerging from such improved forms of entropy and check if they correctly reproduce known deviations determined via non entropic arguments (\textit{theoretical check});
\item[(c)]to open the possibility of a comparison with experimental tests  of Newton's law, which have already been performed \cite{newton} and, more importantly, might be performed in the future \cite{newnewton} (\textit{experimental check}). 
\end{itemize}
In the following sections we will consider two cases in which there are modified forms of holographic entropies, i.e. noncommutative gravity and ungravity.  The point c) will become more interesting in the last section. When deviations are not known \textit{a priori} by means of other approaches, we might think to exploit Verlinde's procedure starting from a generic form of the entropy in view of a forthcoming experimental check. Conversely we could start from already established Newton's law deviations to determine the form of the entropy and pick up the correct microscopic/quantum theory of gravity.

\section{Microscopic description of gravity} As mentioned above, entropy connects the standard description of gravity with the underlying microstructure of a quantum spacetime. Therefore if we can determine  an appropriate form of the entropy, even in terms of some effective microscopic degrees of freedom, we can expect to provide the correct description of the underlying microstructure. With this in mind, we consider 
\be
\Delta S_{\Omega} =k_B\Delta A\left(\frac{c^3}{4\hbar G}+\frac{\partial s}{\partial A}\right)
\label{genentropy}
\ee
as a generic change of entropy modification. Here $s(A)$, a function of the area, encodes deviations. 
Without losing in generality, but having in mind noncommutative geometry as a specific tool for the description of the microscopic structure of a quantum manifold, we start a revision of Verlinde's assumptions. Noncommutative geometry encodes  spacetime microscopic degrees of freedom by means of a new uncertainty relation among coordinates
\begin{equation}
\Delta x^\mu \Delta x^\nu\ge \theta.
\end{equation}
The parameter $\theta$ has the dimension of a length squared and emerges as a natural ultraviolet cutoff from the geometry when coordinate operators fail to commute
\begin{equation}
[x^\mu, x^\nu]=i\Theta^{\mu\nu}
\label{nc}
\end{equation}
with $\theta=|\Theta^{\mu\nu}|$.  In other words, the spacetime turns out to be endowed with an effective minimal length beyond which no further coordinate resolution is possible.  This a feature of the phenomenology of any approach to quantum gravity and it can be found not only in noncommutative geometry (for reviews see \cite{review}), but also in the framework of loop quantum gravity, generalized uncertainty principle, asymptotically safe gravity etc. The scale at which the minimal length emerges is not specified \textit{a priori}, and it is kept generic saying that at the most $\sqrt{\theta}<10^{-16}$ cm, namely, smaller than the typical scale of the standard model of particle physics.  Along this line of reasoning, we have to revise at least two of Verlinde's assumptions in order to introduce the noncommutative scale $\sqrt{\theta}$. For generality we keep $\sqrt{\theta}\neq \ell_P$. The new scheme now reads:
\begin{list}{}{}
\item[1 bis)] Because of the presence of an uncertainty on $\Omega$, there exists a fundamental unit $\Delta S_{\theta}$, which is perceived at the displacement $\Delta x_{min}\propto \lambda_m$. Therefore the change of entropy is 
\begin{equation}
\Delta S_\Omega=\Delta S_\theta\left(\frac{\Delta x}{\Delta x_{min}}\right)
\end{equation}
where for later convenience we set $\Delta x_{min}=\frac{\alpha^2}{8\pi}\lambda_m$.
\item[3 bis)] On the surface $\Omega$, the fundamental unit of surface is determined by the microscopic theory and coincides with $\theta$. Therefore the number of bits reads
\begin{equation}
N=\frac{A_\Omega}{\theta}.
\end{equation}
\end{list}
As we shall see below the introduction of the parameter $\alpha$ lets us keep the noncommutative and the Planck scale distinct. 
From the above new assumptions and from 
\begin{equation}
\Delta S_{\theta}=k_B\theta\left(\frac{c^3}{4\hbar G}+\frac{\partial s}{\partial A}\right)
\end{equation}
one obtains the temperature
\begin{equation}
T=\frac{M}{r^2}\frac{\theta c^2}{2\pi k_B}
\label{temp2}
\end{equation}
and 
\begin{equation}
F=\frac{Mm}{r^2}\left(\frac{4 c^3 \theta^2}{\hbar\alpha^2}\right)\left[\frac{c^3}{4\hbar G}+\frac{\partial s}{\partial A}\right].
\label{ncforce}
\end{equation}
Equation (\ref{temp2}) shows that the temperature is proportional to $\theta$. This let us interpret the temperature as a macroscopic effect of  quantum mechanical fluctuations of the noncommutative manifold. Furthermore, this implies that, for the third law of thermodynamics, it is impossible to reach the limit $\theta\to 0$ for the commutator in (\ref{nc}). In other words, the description of  spacetime in terms of a smooth differential manifold would be equivalent to a thermodynamical system at the absolute zero. 
We can therefore conclude that the spacetime can never escape its noncommutative/thermal state (except the asymptotic limit $r\to\infty$).

Equation (\ref{ncforce}) is the entropic force, which coincides with  Newton's law to first order if $\theta=\alpha\ell_P^2$. As a result the modified Newton's law reads
\begin{equation}
F=\frac{GMm}{r^2}\left[1+4\ell_P^2\frac{\partial s}{\partial A}\right].
\label{Newtoncorr}
\end{equation}
Up to now we have not yet invoked a specific form of the entropy coming from noncommutative geometry. We simply
showed the consequences of the occurrence of noncommutative effects at a scale different from the Planck scale.
 
\subsection{Noncommutative geometry corrections to Newton's law}
We face now the problem of implementing a specific form for the area/entropy relation in noncommutative geometry. To this purpose we consider just a particular approach to noncommutative geometry in which a minimal length is implemented in spacetime by averaging coordinate operator fluctuations \cite{NCQFT}. This way of implementing a minimal length could sound ``minimalistic,'' since most of the noncommutative character of the manifold is actually lost or ``averaged out.'' However, this approach turns out to be very useful for obtaining in a transparent way primary effects emerging from noncommutative geometry (for other formulations of noncommutative gravity see, for instance, \cite{Vassilevich:2009pp}). In addition, since our analysis is devoted to quantum mechanical corrections to a classical law, we do not need anything more than such primary effects. We recall now that averaging noncommutative coordinate operators leads to an effective delocalization of any pointlike source. As a result new regular black hole solutions have been derived solving Einstein equations with a modified, i.e. delocalized, source term \cite{ncbhs}. A lot of work has been devoted to the understanding of the role of these black holes in the context of TeV gravity at the LHC \cite{NCBHpheno}. Furthermore, the new black hole thermodynamics was found to be modified with respect to the conventional scenario: In place of a runaway terminal phase of evaporation, these new black holes cool down to a zero temperature black hole remnant configuration, entirely governed by microscopic fluctuations of the manifold, encoded in the parameter $\theta$ \cite{NCthermo}.  More specifically for the Schwarzschild case the Hawking temperature reads 
\be
T_H=\frac{\hbar c}{4\pi k_B r_H}\left[1-\frac{r_H^3}{4\theta^{3/2}}\frac{e^{-r_H^2/4\theta}}{\gamma(3/2; r_H^2/4\theta)}\right]\ 
\label{nctemp}
\ee
where
\begin{equation}
\gamma(3/2; r_H^2/4\theta)=\int_0^{r_H^2/4\theta}dt\ t^{1/2}e^{-t}
\end{equation}
is the incomplete lower Euler function, while $\Gamma(3/2)=\sqrt{\pi}/2$. We notice that the effects of noncommutative fluctuations are relevant at small scales with respect to $\sqrt\theta$, while at large scales the temperature asymptotically approaches the conventional Hawking result. The entropy of the system can be calculated via
\begin{equation}
dS=\frac{1}{T_H}\ dU(r_H)
\end{equation}
where the thermodynamic energy is nothing but the mass of the black hole as a function of the event horizon, namely,
\begin{equation}
U(r_H)=r_H\ \frac{c^4  }{2G}\left(\frac{\Gamma(3/2)}{\gamma(3/2; r_H^2/4\theta)}\right),
\end{equation}
where $\Gamma(3/2)=\sqrt{\pi}/2$.
From the above relations one obtains
\be
d S=d \left(4\pi r_H^2\right)\ \frac{k_B c 3}{4\hbar G}\ \left(\frac{\Gamma(3/2)}{\gamma(3/2; r_H^2/4\theta)}\right). 
\label{ncentropy}
\ee
We notice that  the noncommutative geometry corrections do not affect the general form of the entropy, apart from a multiplicative term in round brackets. Such a term reduces to $1$ in the classical theory. Therefore, following Verlinde's arguments, we just need to multiply the classical value $\Delta S_\Omega$ to obtain the noncommutative geometry corrected entropy for any holographic surface $\Omega$. As a result, by means of  (\ref{genentropy}) and (\ref{Newtoncorr}), we find in a transparent way and without approximations
\begin{equation}
F=\frac{GMm}{r^2}\left[1+\frac{\Gamma(3/2; r^2/(4\pi\theta))}{\gamma(3/2; r^2/(4\pi\theta))}\right],
\end{equation}
where $\Gamma(3/2; r^2/(4\pi\theta))$ is the incomplete upper Euler function. If we consider only primary corrections in the limit $r\gg\sqrt{\theta}$ we have
\begin{equation}
F\simeq\frac{GMm}{r^2}\left[1+\frac{r\ e^{-r^2/(4\alpha\ell_P^2)}}{\sqrt{\pi\alpha}\ell_P}\ +\frac{r^2 e^{-r^2/(2\alpha\ell_P^2)}}{2\alpha\sqrt{\pi}\ell_P^2}\right].
\end{equation}
We stress that the above result has been obtained from the mere thermodynamical properties of the black hole, properties that we simply have extended to a generic holographic screen in the spirit of Verlinde's procedure.

\subsection{Ungraviton corrections to Newton's law}
Ungravity is the gravitational force, due to the exchange of ungravitons. Unparticles are conjectured particles which extend the concept of the neutrino. Indeed like neutrinos unparticles are supposed to be weakly interacting with standard model particles. Furthermore, like neutrinos, unparticles are described by conformal invariant fields. The novelty, and the difference with respect to neutrinos, is that unparticles are supposed to be massive. To preserve the conformal invariance, an unfield must be endowed with a continuous mass spectrum, labeled by a scaling parameter $d_U$ \cite{unparticle}. In general $d_U$ is a generic real number, but it is often chosen $1<d_U<2$. The case $d_U=1$ corresponds to the absence of effects from unparticle physics. Recently unparticle physics has been applied to gravity. By effectively modifying the Einstein-Hilbert action with an unparticle term, the un-Schwarzschild black hole has been derived as an extension of the Schwarzschild geometry when gravity is mediated by ungravitons, rather than by gravitons \cite{uneuro}. The presence of the scaling parameter $d_U$ forces a fractalization of the event horizon of the un-Schwarzschild black hole. As a result the horizon dimension is $2d_U$. This fact has been confirmed by recent studies about the spectral properties of a quantum spacetime \cite{micro}. In particular a new kind of spectral dimension has been proposed, i.e. the unspectral dimension, stemming from the diffusion of a unfield. As a result, it has been shown that unparticles play a crucial role in determining the actual dimension perceived by a random walker in the trans-Planckian regime. The thermodynamic properties of the un-Schwarzschild black hole have been determined too. In particular the Hawking temperature reads
\begin{equation}
T=\frac{\hbar c}{4\pi k_B r_H}\left[1+\frac{2(2d_U-1) \Gamma_U}{1+\Gamma_U (\frac{R}{r_H})^{2d_U-2}}\left(\frac{R}{r_H}\right)^{2d_U-2}\right]
\end{equation} 
where 
\begin{equation}
\Gamma_U=\frac{2}{\pi^{2d_U-1}}\frac{\Gamma(d_U-1/2)\Gamma(d_U+1/2)}{\Gamma(2d_U)}
\end{equation}
and 
\begin{equation}
R=\frac{1}{\Lambda_U}\left(\frac{M_P}{M_U}\right)^{1/(d_U-1)}.
\end{equation}
Here $M_U$ is the ungravity coupling constant, while $\Lambda_U$ is the critical energy scale below which scale invariant properties of unparticles would emerge. The thermodynamic energy of the system, namely, the black hole mass as a function of the horizon reads
\begin{equation}
U(r_H)=r_H\ \frac{c^4  }{2G}\left(\frac{1}{1+\Gamma_U (\frac{R}{r_H})^{2d_U-2}}\right).
\end{equation}
From the above relations one obtains
\be
d S=d \left(4\pi r_H^2\right)\ \frac{k_B c^3}{4\hbar G}\ \left(\frac{1}{1+\Gamma_U (\frac{R}{r_H})^{2d_U-2}}\right). 
\label{unentropy}
\ee
Once integrated, this relation leads, in the ungravity dominated phase, to the aforementioned complete fractalization of the horizon, i.e.,
\begin{equation}
S=\frac{k_B c^3}{\hbar G}\frac{\pi R^{2-2d_U}}{d_U \Gamma_U}\ r_H^{2d_U}.
\end{equation}
Since we have not made any assumption about the noncommutative nature of spacetime we can keep Verlinde's original assumptions or equivalently $\alpha=1$. We notice that (\ref{unentropy}) is equivalent to the conventional form of the entropy, apart a multiplicative factor which encodes ungraviton corrections. Therefore, by extending this relation to any arbitrary holographic screen, we find that the entropic unforce reads
\begin{equation}
F=\frac{GMm}{r^2}\left[1-\frac{\Gamma_U (\frac{R}{r_H})^{2d_U-2}}{1+\Gamma_U (\frac{R}{r_H})^{2d_U-2}}\right]
\end{equation}
that, in the ``weak unfield limit'', becomes
\begin{eqnarray}
&&F\simeq \\ &&\frac{GMm}{r^2}\left[1-\Gamma_U \left(\frac{R}{r_H}\right)^{2d_U-2}+\left(\Gamma_U \left(\frac{R}{r_H}\right)^{2d_U-2}\right)^2\right].\nonumber
\end{eqnarray}

\section{Final remarks}
Verlinde has extended the validity of the holographic principle to a generic screen. We have used this extension to derive the corrections to Newton's law in both noncommutative gravity and ungravity. In particular, we have also given a new thermodynamical picture of noncommutativity. A noncommutative manifold is equivalent to a thermodynamic system, whose temperature is proportional to the noncommutative parameter $\theta$. The commutative limit is therefore forbidden by thermodynamics arguments, since it would correspond to a zero temperature state, which can be reached only asymptotically. As a theoretical check for the consistency of the procedure, we have to  compare our results with the already known Newton's law deviation emerging both from noncomutative geometry \cite{Gruppuso} \footnote{Corrections to  Newton's potential due to noncommutative effects have also been derived in \cite{Harikumar:2006xf} within another approach to noncommutative gravity.} and ungravity \cite{uneuro}. 
Within some approximations, the deviations coincide with our results. This by itself sheds light about the Verlinde's derivation, making his procedure more robust, not confined to the mere case of classical physics and not interpretable as a smart dimensional manipulation. In particular we support the Verlinde's crucial assumption about the existence of an entropy associated to a surface even in the absence of horizons. This conclusion is in agreement to what found in the framework of loop quantum gravity, where the Verlinde's approach again matches  known results \cite{vloop}. In addition, with the eventual experimental proof of one of the deviations considered, we would have the double result of confirming the noncommutative/ungravity nature of the corrections as well as the entropic interpretation of these effects. 
Anyway, there is something more. 

We have followed the scheme ``modified theory of gravity'' $\to$ ``modified black hole entropy'' $\to$ ``modified holographic surface entropy'' $\to$ ``Newton's law corrections.'' 
We have seen that at the very end  both noncommutative geometry and ungravity corrections to Newton's law emerge from a modified form of the gravitational constant $G$. Indeed in both (\ref{ncentropy}) and (\ref{unentropy}), the constant $G$ is accompanied by a multiplicative factor which resumes the effects of the theory under investigation. The same factor eventually provides the Newton's law deviations. We intend to show that this is not a specific property of the cases we considered.
With hindsight we could suppose to extend our procedure to a more generic situation. We start from a black hole metric like in (\ref{metric}) and we assume that
\begin{equation}
f(r)=1-\frac{2M}{r^{n-2}c^2}{\cal G}(r)
\label{nm}
\end{equation} 
where $n$ is the number of spatial dimensions and ${\cal G}$ incorporates some effects, like noncommutativity, ungravity, asymptotically safe gravity, extra dimensions, electrostatic energy, etc. The only requirement is that in the limit for which the theory approaches the classical, neutral case one must recover ${\cal G}\to G_{(n)}$, where $G_{(n)}$ is the gravitational constant in $n+1$ dimensions. Suppose we now study the energy $U=c^2 M$ which satisfies the law of thermodynamics
\begin{equation}
dU(r_H)=T dS+\Phi(r_H)dq
\end{equation}
where $\Phi(r_H)$ is the electrostatic potential on the event horizon. The thermodynamic energy is in general a function of both $r_H$ and $q$ and reads
\begin{equation}
 dU=\frac{\partial U}{\partial r_H}dr_H + \frac{\partial U}{\partial q}dq.
\end{equation}
From the above expressions we get
\begin{equation}
dS=\frac{1}{T}\frac{\partial U}{\partial r_H}dr_H.
\end{equation}
Therefore, for (\ref{nm}) we have that
\begin{equation}
U(r_H)=\frac{r_H^{n-2}c^4}{2 {\cal G}}
\end{equation}
and
\begin{equation}
\frac{\partial U}{\partial r_H}=\frac{r_H^{n-3}c^4}{2 {\cal G}}\left[(n-2)-r_H\frac{{\cal G}^\prime}{{\cal G}}\right].
\end{equation}
On the other hand the temperature reads
\begin{equation}
T=\frac{\hbar c}{4\pi k_B r_H}\left[(n-2)-r_H\frac{{\cal G}^\prime}{{\cal G}}\right]
\end{equation}
and therefore we obtain a general expression for the entropy as
\begin{equation}
dS=\frac{(8\pi)k_B c^3 r_H^{n-2}}{4\hbar {\cal G}}\ dr_H.
\end{equation}
We notice that the function ${\cal G}$ enters naturally as a correction of the entropy too. The above formula can be written as
\be
dS=\frac{1}{(n-1)}\ \frac{ \Gamma\left(\frac{n}{2}\right)}{ \pi^{\frac{n}{2}-1}}\ \frac{k_B}{{\cal G}}\ dA_{(n-1)},
\ee
which is formally the conventional entropy apart from ${\cal G}$ in place of $G_{(n)}$.
Here we have used the natural units $\hbar=c=1$ and $A_{(n-1)}$ is the $(n-1)$-sphere surface area, namely,
\be
A_{(n-1)}=n\ \frac{\pi^{n/2}}{\Gamma(\frac{n}{2} + 1)}\ r^{n-1}.
\ee
Given this point, if we believe in Verlinde's arguments about the extension of this form of entropy to any holographic  surface, we have just to replace the $G_{(n)}$ with ${\cal G}$. Before proceeding, we have to take into account that, like for the noncommutative geometry case, the effects enclosed in ${\cal G}$ occur at an energy scale ${\cal M}$, which is in general different from the fundamental scale $M_\star$, i.e., $G_{(n)}=M_\star^{1-n}$. Therefore we will have $\alpha\neq 1$ and the fundamental unit of entropy reads
\begin{equation}
dS_{{\cal M}}=k_B \ {\cal M}^{1-n}\left[\frac{1}{M_\star^{1-n}}\frac{\Gamma\left(\frac{n}{2}\right)}{(n-1)\pi^{\frac{n}{2}-1}}+\frac{\partial s}{\partial A_{(n-1)}}\right].
\end{equation}
At the scale ${\cal M}$,  the fundamental unit of surface also changes and it is ${\cal M}^{1-n}$. Then the number of bits reads
\begin{equation}
N=\frac{A_{(n-1)}}{{\cal M}^{1-n}}.
\end{equation}
As a result we find that the entropic force is
\begin{equation}
F=\frac{G_{(n)}Mm}{r^{n-1}}\left[1-\frac{\Delta G_{(n)}}{{\cal G}}\right]
\end{equation}
where $\Delta G_{(n)}={\cal G}-G_{(n)}$ accounts for deviations from the classical case. Here we have chosen 
\begin{equation}
\alpha^2=\frac{8}{\pi^{n-2}}\frac{\Gamma^2\left(\frac{n}{2}\right)}{(n-1)}\left(\frac{{\cal M}}{M_\star}\right)^{2-2n}
\end{equation} 
to match the Newton's law to the first order.
Now we can use the above general result for some purposes. First, it is no longer necessary to repeat the above lengthy calculations every time. For every form of ${\cal G}$, we can derive Newton's law by entropic arguments automatically. Second, if for some reasons we just know the form of ${\cal G}$ or the form of entropy but we ignore or we cannot perform the weak field limit, we could use the generality of the above formula to get deviations to Newton's law; conversely, if we start from known (or future) experimental observations of Newton's law deviations we can go back and pick the correct form of the entropy as well as ${\cal G}$. This, in principle, would let us get information about the correct formulation of quantum gravity.
For these reasons, we strongly believe that the relation between entropy and Newton's law will deserve further investigations.  
\begin{acknowledgments}
\noindent P.N. is supported by the Helmholtz International Center for FAIR within the
framework of the LOEWE program (Landesoffensive zur Entwicklung Wissenschaftlich-\"{O}konomischer Exzellenz) launched by the State of Hesse. P.N. would like to thank the Perimeter Institute for Theoretical Physics, Waterloo, Ontario, Canada for the kind hospitality during the period of work on this project. P.N. warmly thanks L. Modesto for a valuable suggestion. 
\end{acknowledgments}

\end{document}